\documentclass[preprint,reprintnumbers,amsmath,amssymb]{revtex4-1}
\usepackage{amssymb}

\usepackage{graphicx}% Include figure files
\usepackage{dcolumn}% Align table columns on decimal point
\usepackage{bm}% bold math
\usepackage{upgreek}

\usepackage{epstopdf} 
\begin{document}

\title{Plasma-induced magnetic patterning of FePd thin films without and with exchange bias}

\author{Wei-Hsiang Wang}
\author{Po-Chun Chang}
\author{Pei-hsun Jiang}
\altaffiliation{E-mail: pjiang@ntnu.edu.tw}
\author{Wen-Chin Lin}
\altaffiliation{E-mail: wclin@ntnu.edu.tw}
\address{Department of Physics, National Taiwan Normal University, Taipei 116, Taiwan}

\begin{abstract}
We demonstrate control of magnetic domain structures in continuous FePd thin films by patterning their surfaces with plasma treatment. The Fe-oxide layer formed on the surface upon ambient exposure of the FePd alloy thin film grown on an Al$_2$O$_3$(0001) substrate was patterned into microstructures by e-beam lithography followed by O$_2$- or Ar-plasma treatment. Microscopic pinning of magnetic domain walls in the thin films is then observed by magneto-optic Kerr effect microscopy, with the magnetic field needed to reverse the magnetization of the plasma-treated areas being larger than that for the untreated areas. An intriguing competition between the uniaxial anisotropy and the exchange bias is also observed in the system. This study demonstrates that patterning of the film surface with plasma treatment can be an easy and efficient method for sophisticated engineering of magnetic structures in thin films, and therefore has potential application in developing future data-storage and spintronic devices.

\bigskip
\setlength\parindent{0pt} \textit{Keywords:} magnetic patterning, thin films, plasma, exchange bias, magneto-optic Kerr effect microscopy

\end{abstract}

\maketitle

\section{Introduction}

Engineering of local magnetic properties of continuous thin films has long been an important topic in design and fabrication of future spintronic devices, high-density magnetic data-storage devices, and magnetologic devices. Since Chappert \textit{et al.}~first reported the ability of ion irradiation to induce modification of local magnetic properties without altering topographic features,\cite{Chappert1998} ion-induced magnetic patterning has been investigated intensively and demonstrated to be realized through patterned modulations of magnetic anisotropy,\cite{McCord2005, McCord2009, Merkel2008, Jaafar2011} saturation magnetization,\cite{McCord2008,Bali2014} or exchange bias (EB).\cite{Fassbender2008} Most of these modulations rely mainly on changes in the interfacial structures, but also in the case of alloys on changes in the chemical ordering.\cite{Ravelosona2000, Menendez2008} Ion irradiation with sufficient energy and fluence has been used in most studies to achieve changes in local magnetic properties for magnetic patterning. By contrast, plasma treatment was utilized in very few experiments for the same purpose.\cite{Menendez2010}

Among potential materials for ultrahigh-density magnetic recording media, the FePd ferromagnetic (FM) compound has been investigated intensively in recent years. FePd with a tetragonal \textit{L}1$_0$ phase has attracted much attention in light of its uniaxial magnetic anisotropy (UMA),\cite{Ivanov1973} perpendicular magnetic anisotropy (PMA),\cite{Wei2009} and high resistance to corrosion.\cite{Kryder2008} With particular fabrication methods such as specific deposition geometries or use of buffer layers, UMA and/or weak PMA are found in some FePd and other Fe-based alloys even without long-range \textit{L}1$_0$ ordering.\cite{Chi2012,Clavero2006,Jaafar2011} Capping of proper materials %
such as Pd on FePd has been demonstrated to induce specific interaction at the interface due to the anisotropy of FePd,\cite{Durr1999,Clavero2008} which we consider can be a potential means of magnetic engineering. In this study, we demonstrate an easy method that utilizes e-beam lithography and O$_2$- or Ar-plasma treatment to magnetically pattern an FePd thin film with a surface Fe-oxide layer formed upon ambient exposure. The behaviors of the magnetic microstructures are investigated in detail. It is observed that the plasma treatment promotes further Fe oxidation, which prominently enhances the Pd concentration in the Pd-rich phase\cite{Hsu2017,Cialone2017} formed beneath the Fe oxide %
on the upper side of the FePd layer. This alters the magnetic properties of the system, leading to interesting effects of domain pinning and competition between the UMA and EB observed in magneto-optic Kerr effect (MOKE) microscopy experiments.

\section{Material and methods}

\begin{figure}
	\centering
\includegraphics[width=0.7\textwidth]{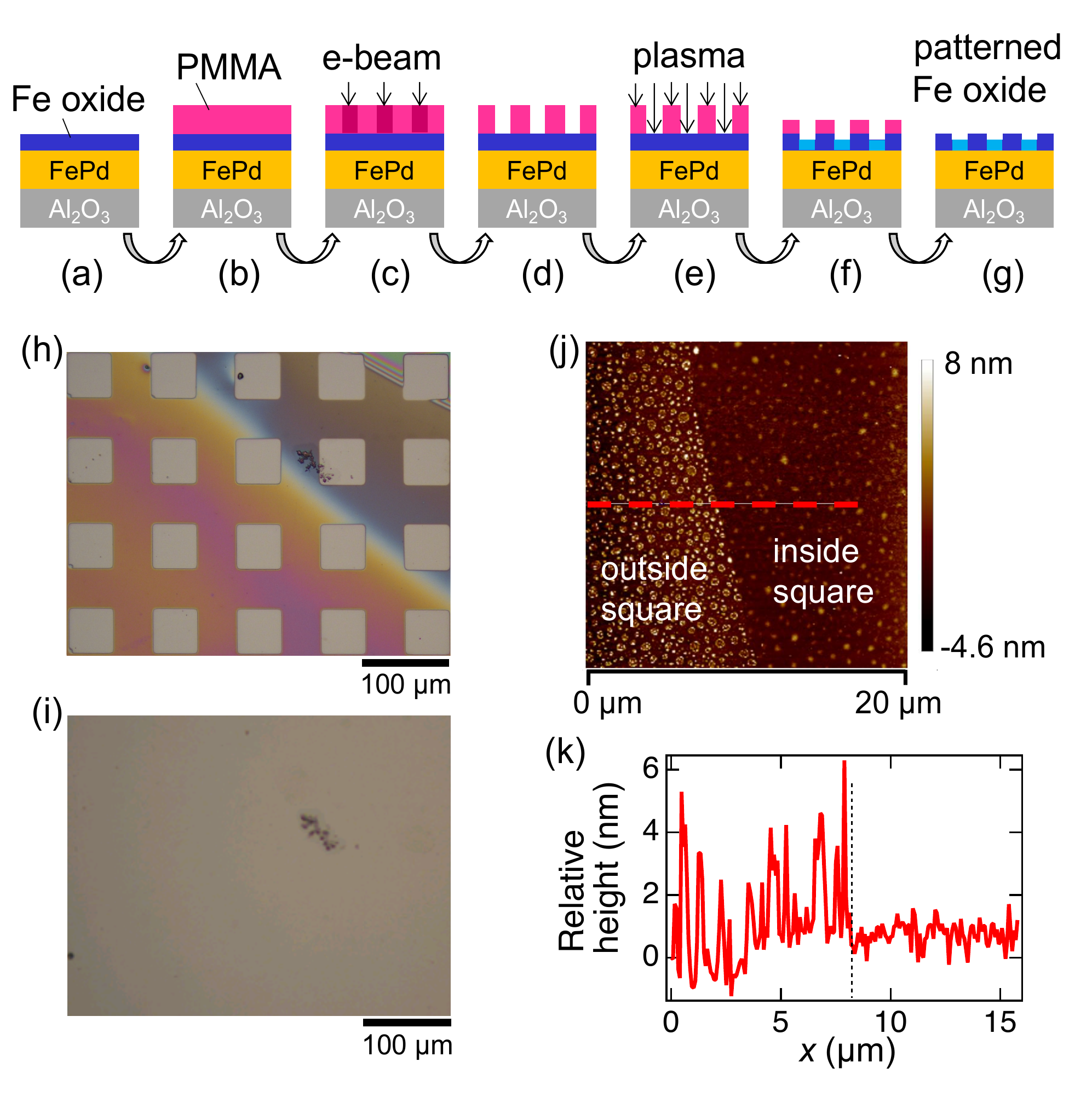}
	\caption{(Color online) (a)--(g) illustrate the surface patterning process including e-beam lithography and plasma treatment. (h) and (i) are the optical images of the Fe oxide/FePd film at the (d) and (g) stages, respectively. (j) exhibits the atomic force microscopy image of the surface-patterned Fe oxide/FePd film, with the dashed-line profile shown in (k). }\label{Fig1}
\end{figure}

The 30-nm-thick FePd (50:50) alloy films were deposited on Al$_2$O$_3$(0001) substrates by e-beam heated co-evaporation in an ultrahigh vacuum chamber with a base pressure of $3\times10^{-9}$ mbar. Two evaporation guns were both aligned at 45$^{\circ}$ to the normal. This oblique deposition geometry allows UMA to be developed on the surface plane.\cite{Chi2012} The alloy compositions and film thicknesses were controlled by the respective deposition rates of the elements, and were calibrated by Auger electron spectroscopy, X-ray photoelectron spectroscopy (XPS), atomic force microscopy, and transmission electron microscopy with energy dispersive X-ray spectroscopy. X-ray diffraction results showed that the films grown do not exhibit long-range \textit{L}1$_0$ ordering. Then, the FePd films were naturally oxidized after exposure to ambient conditions, which lead to formation of a self-assembled thin Fe-oxide layer on the surfaces. E-beam lithography and O$_2$- or Ar-plasma treatment (Harrick Plasma Cleaner PDC-32G, radio frequency 13.56 MHz) were utilized for the patterning process on the surface of Fe oxide/FePd. The surface morphologies and chemical compositions are examined by atomic force microscopy and XPS, respectively. The microscopic magnetic behaviors, including magnetic hysteresis loops and magnetic domain structures, are monitored by a MOKE microscope with the presence of an in-plane magnetic field $H$.\cite{Chang2018}

As shown in Figs.~\ref{Fig1}(a)--1(g), e-beam lithography was performed to create a microstructure pattern in poly(methyl methacrylate) (PMMA) for selective plasma treatment of the film surface afterwards, which results in a patterned Fe-oxide layer on top of the FePd alloy film. The optical images of an Fe oxide/FePd film at stages of Figs.~\ref{Fig1}(d) and 1(g) are shown in Figs.~\ref{Fig1}(h) and 1(i), respectively. Fig.~\ref{Fig1}(h) demonstrates a patterned PMMA layer with arrays of $50 \times 50$ $\upmu$m$^2$ square holes spaced 50 ${\upmu}$m apart. At a base pressure of 200 mTorr of the plasma chamber, 3 minutes of 10.5-Watt O$_2$ plasma was used to bombard the square areas, leaving the areas under PMMA intact. The plasma effect on the squares is hardly observable in the optical image in Fig.~\ref{Fig1}(i), but manifests itself in the atomic force microscopy image in Fig.~\ref{Fig1}(j), with the dashed-line profile shown in Fig.~\ref{Fig1}(k).
In the intact region outside the squares, the FePd surface is still covered by the Fe-oxide clusters naturally self-assembled after ambient exposure. Inside the squares, however, the Fe-oxide layer has become much smoother with reduced grain sizes after the O$_2$ plasma treatment, and has been identified by XPS depth profiling technique to be only $\sim$1 nm thick.

\section{Experimental results and discussion}
\subsection{Magnetic patterning with O$_2$- or Ar-plasma treatment}

\begin{figure*}
	\centering
\includegraphics[width=1\textwidth]{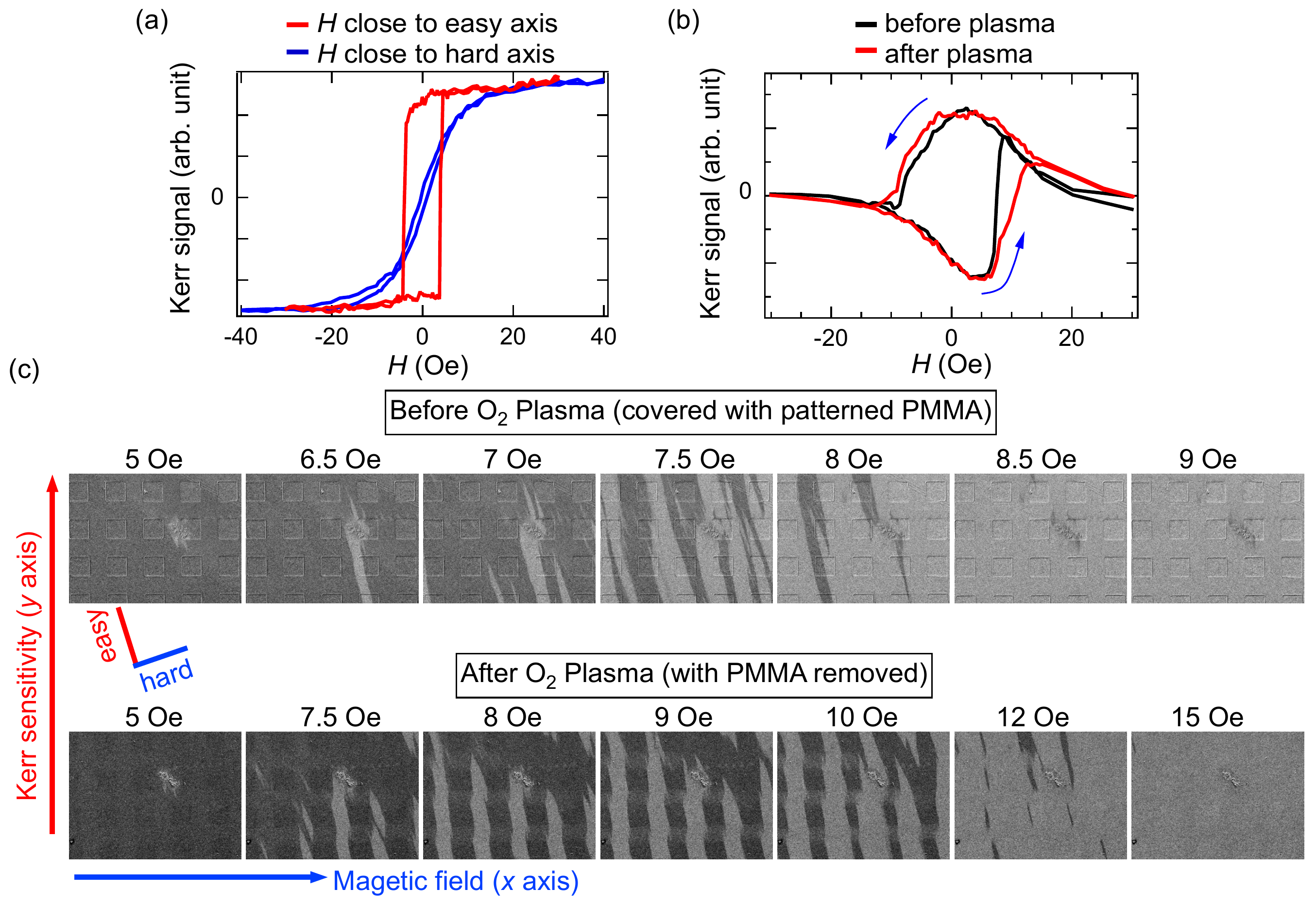}
	\caption{(Color online) (a) \textit{Longitudinal} MOKE hysteresis loops of the surface-patterned Fe oxide/FePd film with the magnetic field along the directions close to the magnetic easy (red curve) and hard (blue curve) axes, respectively. (b) \textit{Transverse} MOKE hysteresis loops of the Fe oxide/FePd film with the magnetic field along the direction close to the hard axis before (black curve) and after (red curve) plasma treatment, respectively. (c) Evolutions of the magnetic domain structures simultaneously recorded with the hysteresis loops in (b). Each image size is $450 \times 350$ $\upmu$m$^2$.}\label{Fig2}
\end{figure*}

The FePd alloy films reveal UMA induced by the oblique co-deposition. Fig.~\ref{Fig2}(a) shows the MOKE hysteresis loops of a surface-patterned Fe oxide/FePd film measured in the \textit{longitudinal} geometry with the magnetic field applied along the directions close to the magnetic easy (red curve) and hard (blue curve) axes, respectively. (The direction of the applied field or the Kerr sensitivity throughout the experiment is at an angle of $\sim$$15^{\circ}$ to the hard or easy axis, and these directions are labeled in the figures as $x$ and $y$ axes, respectively.) When the field direction is close to the easy axis, a square hysteresis loop is observed with a magnetic coercivity $H_\mathrm{c}$ of $\sim$4 Oe. When the field is instead applied in a direction close to the hard axis, which is in the film plane orthogonal to the easy axis, a slim hysteresis loop with extremely small remanence is observed. Fig.~\ref{Fig2}(b) shows the MOKE hysteresis loops measured with the magnetic field applied along the direction close to the hard axis in the \textit{transverse} geometry, in which the plane of incidence is orthogonal to the magnetic field. The black and red curves represent the transverse MOKE hysteresis loops measured from the same region of the film before (Fig.~\ref{Fig1}(d)) and after (Fig.~\ref{Fig1}(g)) plasma treatment, respectively. The hysteresis loop becomes wider after plasma treatment. The asymmetry of the hysteresis loops may originate from imperfect alignment between the field and the magnetic structure with higher-order anisotropies in a heterostructure with complexity.\cite{Camarero2005} Displayed in Fig.~\ref{Fig2}(c) is the evolution of the magnetic domain structures simultaneously recorded with the hysteresis loops in Fig.~\ref{Fig2}(b) with increasing magnetic field, and the effect of surface patterning can be clearly seen in these Kerr images. Both series of the images before and after plasma treatment show the preferred alignment of the magnetic domains along the easy axis. However, the evolution patterns are quite different from each other. Before plasma treatment with the patterned PMMA layer on the surface, there is no clear correlation between the magnetic domain structure and the arrays of $50 \times 50$ $\upmu$m$^2$ squares in PMMA. This indicates that the PMMA capping does not induce any observable change in the magnetic properties of the Fe oxide/FePd film, and thus the magnetic domain structure is irregular. By contrast, after plasma treatment and PMMA removal, a magnetic domain structure with periodic stripes is observed. During the reversal of the magnetization of the film from the negative (dark gray) to the positive (light gray) direction, the magnetic domains in the plasma-treated squares seem to be pinned, presenting a delay in their magnetization reversal, until a larger magnetic field is applied. Therefore, the field ($\sim$15 Oe) required for the film after patterned plasma treatment to complete the magnetization reversal is higher than that ($\sim$9 Oe) before plasma treatment. This is consistent with the observation in Fig.~\ref{Fig2}(b) with the after-plasma hysteresis loop being wider than the before-plasma loop.

\begin{figure}
	\centering
\includegraphics[width=0.65\textwidth]{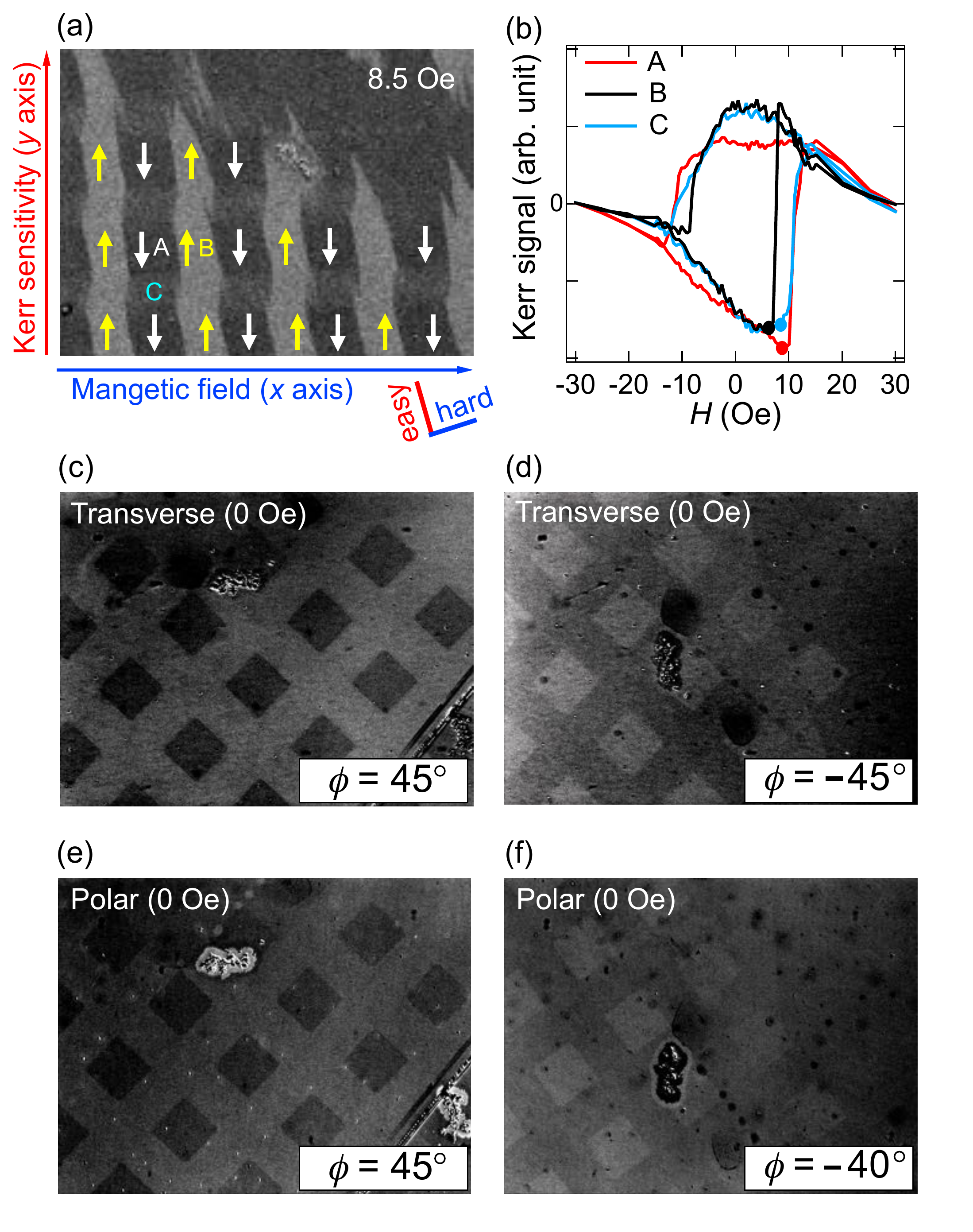}
	\caption{(Color online) (a) Magnetic domain structure of the surface-patterned Fe oxide/FePd film at $H=8.5$ Oe. The arrows indicate the antiparallel magnetization directions of the stripe domains. (b) Spatially-resolved transverse MOKE hysteresis loops from areas A (plasma-treated), B (intact), and C (intact between squares along the easy axis) indicated in (a), respectively.
(c) \& (d) Zero-field transverse MOKE images at $\phi=45^\circ$ and $-45^\circ$, respectively, where $\phi$ is the angle between the current plane of incidence and that in (a). (e) \& (f) Zero-field polar MOKE images at $\phi=45^\circ$ and $-40^\circ$, respectively. Contrast has been enhanced in (c)--(f) for clarity.}\label{Fig3}
\end{figure}

Fig.~\ref{Fig3} illustrates the detailed orientations of the magnetic moments of the surface-patterned Fe oxide/FePd film at $H=8.5$ Oe. In the middle of the magnetization reversal when the field is around 8.5 Oe, the magnetic moments still prefer to align along the easy axis instead of along the direction of the magnetic field, but are constructed into an antiparallel configuration with a geometry matching the columns of the patterned square arrays. The plasma treatment inside the squares induces remarkable pinning of magnetic domains. The spatially-resolved transverse MOKE hysteresis loops of the plasma-treated area A and that of the intact area B as indicated in Fig.~\ref{Fig3}(a) are plotted in Fig.~\ref{Fig3}(b) for comparison, together with the hysteresis loop of the intact area C between the squares along the easy axis. The magnetic moments in areas A and C start to flip up gradually (i.e., to turn from whole to partial dark gray) at a field of around 8.5 Oe, whereas those in area B start to flip up earlier at a lower field around 6.5 Oe. These fields are indicated on the respective hysteresis loops in Fig.~\ref{Fig3}(b) with solid dots, located at the the sharp turning points in the lowest parts of the loop curves. This again suggests that the domain pinning effect is induced by plasma treatment in area A. The magnetic domain in area C between the squares, on the other hand, strongly depends on the anisotropy of the neighboring squares because of lateral modulation along the easy axis.\cite{Jaafar2011,Hierro-Rodriguez2012,Li2002,Menendez2017} Therefore, the field required for area C to reverse the magnetization is about the same as that for area A, which results in dark stripes connecting the squares along the easy axis in the Kerr image.

Exhibited in Figs.~\ref{Fig3}(c) and \ref{Fig3}(d) is the observable reversal of the contrast in the transverse MOKE signal between inside and outside the squares at zero field as the plane of incidence changes from $\phi=45^\circ$ to $-45^\circ$, where $\phi$ is the angle of the present plane of incidence to that (i.e., the $y$ axis) in Fig.~\ref{Fig3}(a). (Contrast has been enhanced in Figs.~\ref{Fig3}(c)--\ref{Fig3}(f) for clarity.) This indicates that plasma treatment truly modifies the magnetic anisotropy of the film, not just its surface morphology, and the easy axis of the new anisotropy may be at an angle of $\sim$$-45^\circ$ to the $y$ axis in Fig.~\ref{Fig3}(a), i.e., $\sim$$-30^\circ$ to the original easy axis. Polar MOKE images at different angles are also obtained as shown in Figs.~\ref{Fig3}(e) and \ref{Fig3}(f), which interestingly reveal perpendicular magnetization of the film and again the contrast between inside and outside the squares. The perpendicular magnetization can be attributed to the anisotropy of FePd, which is known to be a ferromagnet (FM) with weak PMA.\cite{Durr1999}

\begin{figure*}
	\centering
\includegraphics[width=1\textwidth]{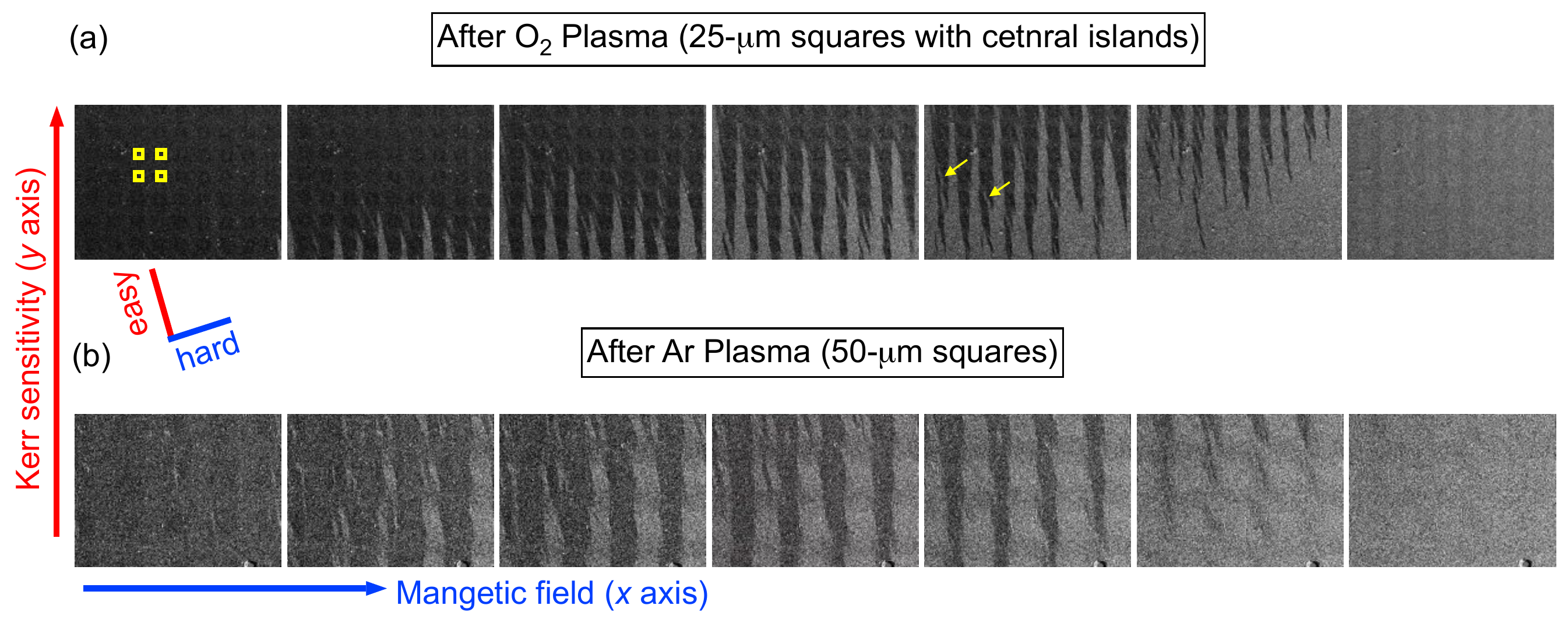}
	\caption{(Color online) Evolutions of the magnetic domain structures as the magnetic field along the direction close to the hard axis is monotonically varied for (a) an O$_2$-plasma treated Fe oxide/FePd film with arrays of $25 \times 25$ $\upmu$m$^2$ squares with central square islands, and for (b) an Ar-plasma treated Fe oxide/FePd film with arrays of $50 \times 50$ $\upmu$m$^2$ squares. Yellow regions in the first image of (a) illustrate some of the plasma-treated areas, with a central $9 \times 9$ $\upmu$m$^2$ square island in each square left untreated under PMMA protection during plasma bombardment. Yellow arrows in the fifth image in (a) indicate magnetization reversals occurring on the central islands.}\label{Fig4a}
\end{figure*}

The structured magnetic domains observed in the surface-patterned Fe oxide/FePd films demonstrate that the method of surface patterning with plasma treatment can modulate the magnetic behaviors of a continuous film and create designed domains at the micro- or potentially even nanoscale. E-beam lithography provides high pattern resolution and excellent design freedom, which allows great flexibility in future developments and applications of magnetic patterning. Shown in Fig.~\ref{Fig4a}(a) is an example of a different pattern composed of smaller squares of sides 25 $\upmu$m, spaced 25 ${\upmu}$m apart. Each square has a central square island of side 9 $\upmu$m that was previously protected by PMMA from O$_2$ plasma. As one can see from the series of images from left to right with increasing magnetic field applied along the direction close to the hard axis, the magnetic patterning and domain pinning phenomena here are similar to those observed in the previous case of $50 \times 50$ $\upmu$m$^2$ squares, except that traits of the central-island effect are found in the present case, with fine light-gray domain strips (some indicated with yellow arrows) running across the square through the central island, showing that the reversal of the magnetization inside a square starts from the \textit{unpinned} central island previously protected from plasma. Fig.~\ref{Fig4a}(b), on the other hand, shows the result of $50 \times 50$ $\upmu$m$^2$ squares treated with Ar plasma instead of O$_2$ plasma, demonstrating that Ar plasma has similar patterning effects to those of O$_2$ plasma. This implies that the sample surface treated with Ar plasma may have undergone structural modification similar to that with O$_2$-plasma treatment.

\subsection{Effect of plasma treatment time on magnetic patterning}

\begin{figure}
	\centering
\includegraphics[width=0.8\textwidth]{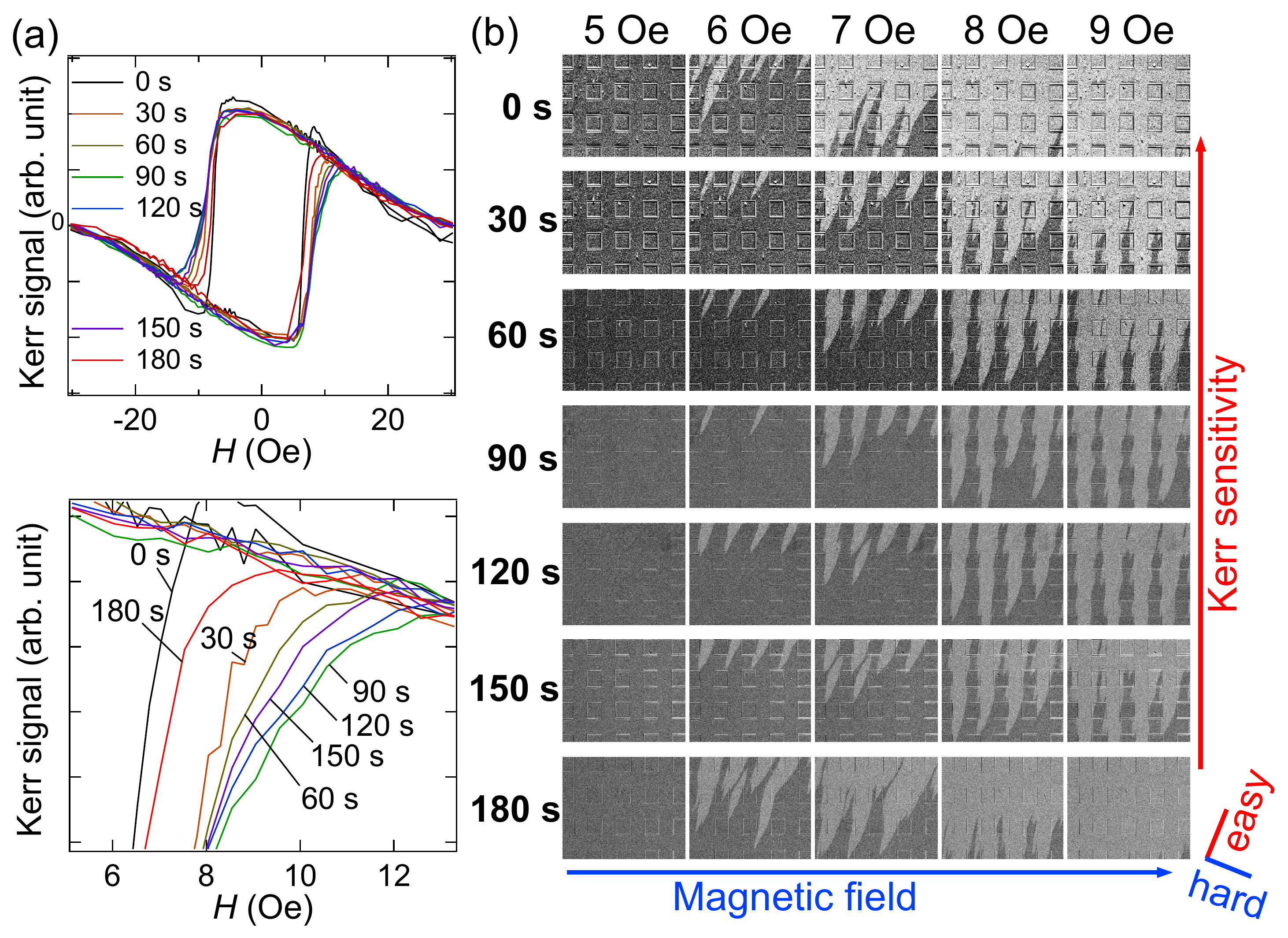}
	\caption{(Color online) (a) Transverse MOKE hysteresis loops of a surface-patterned Fe oxide/FePd film with different treatment times of O$_2$ plasma from 0 to 180 seconds. The lower panel offers a zoomed-in view. (b) Kerr images taken simultaneously with the hysteresis loops in (a) from $H=5$ Oe to 9 Oe. Patterned PMMA is left on the sample surface for easy recognition of the squares.}\label{Fig4}
\end{figure}

The effect of plasma treatment is further examined by studying the evolution of the magnetic domain structure as a function of the treatment time on another Fe oxide/FePd film. Shown in Figs.~\ref{Fig4}(a) and \ref{Fig4}(b) are the transverse MOKE hysteresis loops and the Kerr images after different O$_2$-plasma treatment time durations with the field direction close to the hard axis. The lower panel of Fig.~\ref{Fig4}(a) is a zoomed-in view of the upper panel and focuses on the magnetization reversal process displayed in Fig.~\ref{Fig4}(b). It can be seen that the domain-pinning behavior of this film is most prominent at a treatment time of 90 seconds, from the fact that the periodic dark strips in Fig.~\ref{Fig4}(b) are most persistent at 90 seconds as the field is increased to 9 Oe. As the treatment time is increased further from this optimal value, the domain-pinning effect becomes less pronounced. The hysteresis loops in the lower panel of Fig.~\ref{Fig4}(a) also show that the Kerr signal of the 90-second curve is smaller than all the others when $H\sim 9$ Oe, presenting a most prominent delay in magnetization reversal.

The most probable reason for long-time treatment to fail in domain pinning is that too much plasma bombardment leads to serious interfacial mixing, which is assumed to cause a reduction of the exchange interaction across the interface.\cite{Fassbender2004} As the Fe-oxide layer gets thinner during plasma treatment over a longer duration to provide less protection, the ion bombardment on the FePd layer may cause disorder effects that lead to essential changes in the magnetic properties and exchange interaction.\cite{Fassbender2008,Fassbender2003,Ehresmann2011,Schmidt2014,Bennett2018,Gaul2016} To investigate the possibility of the disorder effects in our case, the penetration depths of plasma ions into the film and the displacement damage profile are estimated by \textsc{srim} simulation.\cite{Ziegler2010} The simulation results confirm that the ions are mostly stopped in the oxide layer of the films treated with the optimal plasma treatment time, making the disorder effects negligible in the FePd layer. The mechanisms involved in the magnetic patterning are most likely to occur mainly in the FePd layer, where a Pd-rich phase is formed upon Fe oxidation, as will be discussed in detail in the following sections.

\subsection{XPS analyses}\label{xps}

\begin{figure}
	\centering
\includegraphics[width=0.7\textwidth]{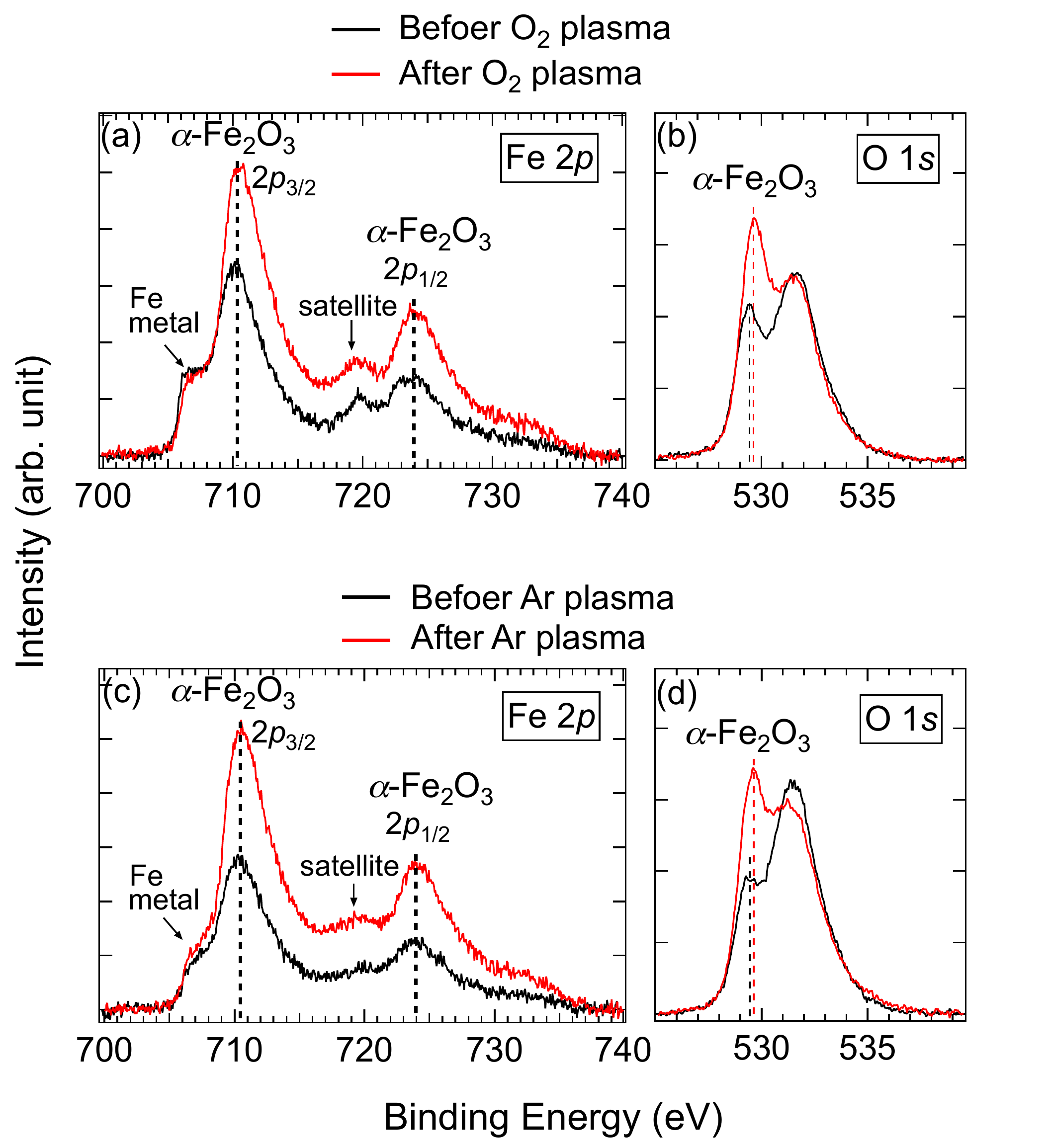}
	\caption{(Color online) (a) Fe 2\textit{p} and (b) O 1\textit{s} XPS of the film surface before and after O$_2$-plasma treatment, respectively. (c) Fe 2\textit{p} and (b) O 1\textit{s} XPS of the film surface before and after Ar-plasma treatment, respectively. $\alpha$-Fe$_2$O$_3$ 2\textit{p} peaks are observed at 710.5 eV and 724.0 eV, and O 1\textit{s} are observed around 529.7 eV.}\label{Fig5}
\end{figure}

The composition of the surface Fe-oxide layer of the film is analyzed using XPS. Displayed in Fig.~\ref{Fig5}(a) are the Fe 2\textit{p} spectra before and after O$_2$-plasma treatment, respectively. It can be seen that, after plasma treatment, the $\alpha$-Fe$_2$O$_3$'s Fe 2\textit{p}$_{3/2}$ peak at 710.5 eV and Fe 2\textit{p}$_{1/2}$ peak at 724.0 eV are both remarkably enhanced. The satellite peak of Fe 2\textit{p}$_{3/2}$ for $\alpha$-Fe$_2$O$_3$ is also observed around 719.5 eV.\cite{Mills1983} Other forms of Fe oxides such as Fe$_3$O$_4$ may coexist in smaller proportions in the oxide layer, if their signature peaks are hidden by the main $\alpha$-Fe$_2$O$_3$ peaks. The Fe metal signal at 707 eV from the FePd underneath the oxide layer can also been seen.   

Presented in Fig.~\ref{Fig5}(b) are the O 1\textit{s} spectra respectively before and after O$_2$-plasma treatment, which also demonstrate an obvious enhancement of the $\alpha$-Fe$_2$O$_3$ signal around 529.7 eV. An additional peak is observed around 531.8 eV, which may be attributed to oxygen vacancies\cite{Wang2019} or surface contaminants of carbon and hydroxide. A very small shift of the $\alpha$-Fe$_2$O$_3$ O 1\textit{s} peak to a higher energy is noticed after plasma treatment, which is probably related to a decrease in the number of oxygen vacancies.\cite{Wang2019} 

\begin{figure}
	\centering
\includegraphics[width=0.45\textwidth]{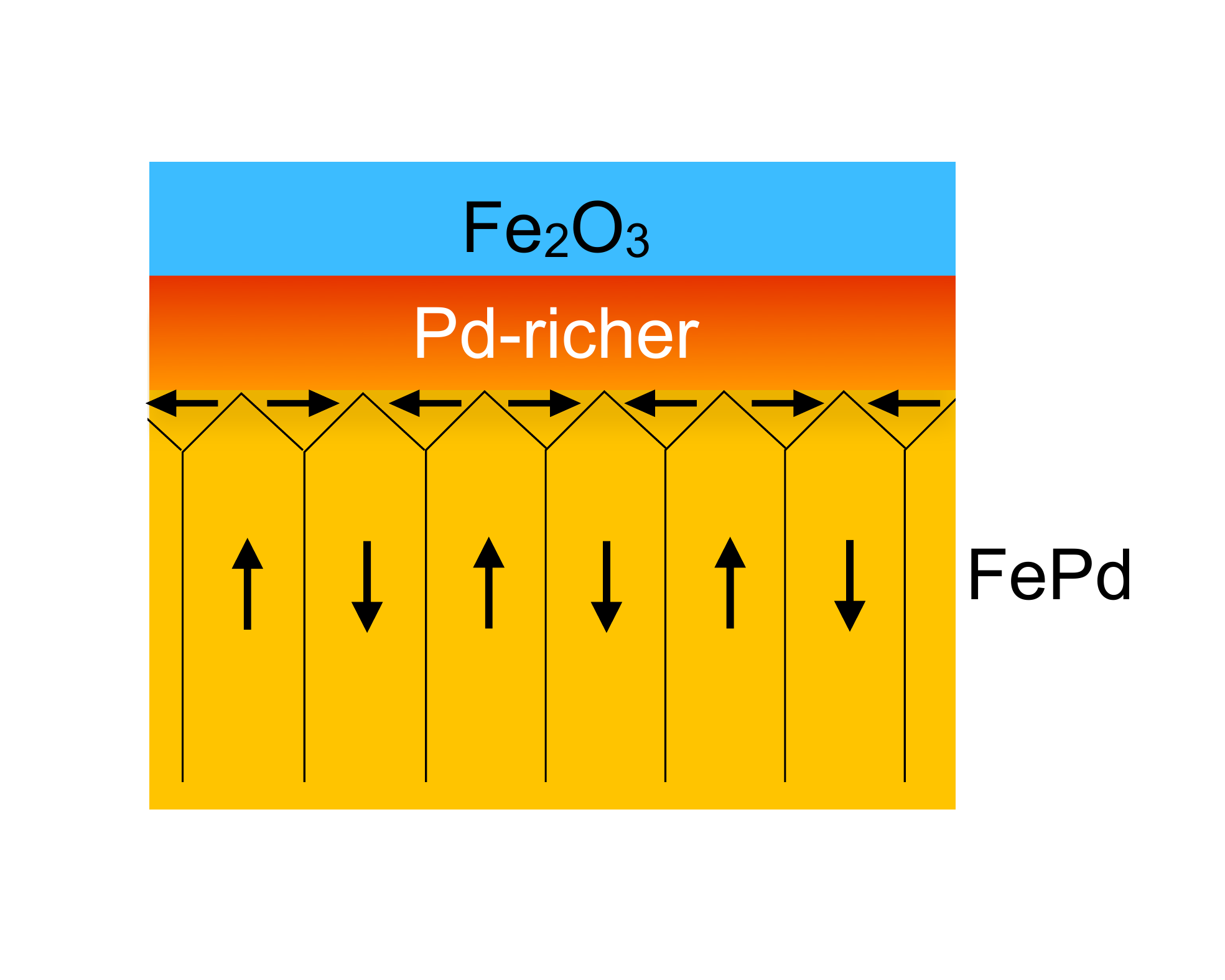}
	\caption{(Color online) Schematic diagram of the Pd-richer/FePd interface underneath the oxide layer in the plasma-treated area. An in-plane anisotropy phase is established near the Pd-richer/FePd interface.}
	\label{Fig_interface}
\end{figure}

These data indicate that the O$_2$-plasma treatment does not remove the oxide layer, but instead modifies the morphology and composition of it, and even further oxidizes more Fe in the existing FePd layer underneath and thus prominently increases the Pd concentration in the Pd-rich phase beneath the oxide\cite{Hsu2017,Cialone2017} (Fig.~\ref{Fig_interface}) with more Fe oxide formed on the surface. The pronounced increase of the $\alpha$-Fe$_2$O$_3$ signal can be ascribed mainly to the oxidation effect of the O$_2$ plasma, and partially to the more-densely-packed finer oxide particles created after plasma treatment and the reduction of surface contaminants identified by other XPS C 1\textit{s} measurements. Similar XPS features are also observed for the sample treated with Ar plasma, as shown in Figs.~\ref{Fig5}(c) and \ref{Fig5}(d). It is interesting to notice that Ar-plasma treatment also results in an enhancement of the oxide signal. The probable explanation is that Ar-plasma treatment gives rise to surface radicals,\cite{Zhang2005} which then promotes oxidation with the presence of water and residual oxygen in the process chamber and on the sample surface.\cite{Surdu-Bob2002}

It is known that local variations in the exchange or magnetocrystalline anisotropy energies can attract or even pin the domain walls under suitable conditions.\cite{Kim2005} Similar phenomena have been intensively studied in antiferromagnet (AFM)/ferromagnet (FM) systems with magnetic defects \cite{Fassbender2008,Kim2005,Nikitenko2000,Vallejo-Fernandez2011} or nonmagnetic dilution treatments.\cite{Vallejo-Fernandez2011,Nowak2002,Fecioru-Morariu2007,Miltenyi2000} With $\alpha$-Fe$_2$O$_3$ being a \textit{canted} AFM at room temperature, it was first speculated that the mechanism of the magnetic patterning might lie in the magnetic heterostructure of the $\alpha$-Fe$_2$O$_3$ layer and the ferromagnetic FePd layer.
However, it requires well-oriented crystal planes (FM coupling within the (0001) planes, and AFM coupling across neighboring planes)\cite{Catti1995} or low temperatures\cite{Zysler1994} for $\alpha$-Fe$_2$O$_3$ to exhibit stable AFM behaviors, and therefore it is unlikely for the amorphous $\alpha$-Fe$_2$O$_3$/FePd interface to host reliable AFM--FM exchange interaction at room temperature.

The most plausible approach to interpret the domain-pinning effect is to consider the interaction at the Pd-rich/FePd interface. FePd is an FM with weak PMA (as revealed in Figs.~\ref{Fig3}(e) and \ref{Fig3}(f)), and Pd capping on FePd has been demonstrated to induce an additional in-plane magnetic anisotropy near the Pd/FePd interface.\cite{Durr1999,Clavero2008} With weak PMA, well-ordered alternating up-and-down magnetic domains\cite{Hierro-Rodriguez2013} are formed in FePd at zero field, and alternating left-and-right in-plane domains  appear near the interface.\cite{Durr1999,Sort2004} As plasma treatment leads to segregation of more Fe atoms from the FePd layer to form more Fe oxide on the surface, the Pd-rich phase\cite{Hsu2017,Cialone2017} near the interface of FePd is turned into a Pd-\textit{richer} phase, establishing a Pd-richer/FePd interface underneath, as illustrated in Fig.~\ref{Fig_interface}. The stronger interfacial in-plane anisotropy created inside the separate plasma-treated squares thereby establishes an anisotropy that is different from the continuous UMA outside the squares, which leads to a delay in reversal of the magnetization observed inside the squares. More theoretical and experimental work may be needed to fully understand the underlying physics.

\subsection{Interplay between exchange bias and uniaxial magnetic anisotropy}

\begin{figure}
	\centering
\includegraphics[width=0.7\textwidth]{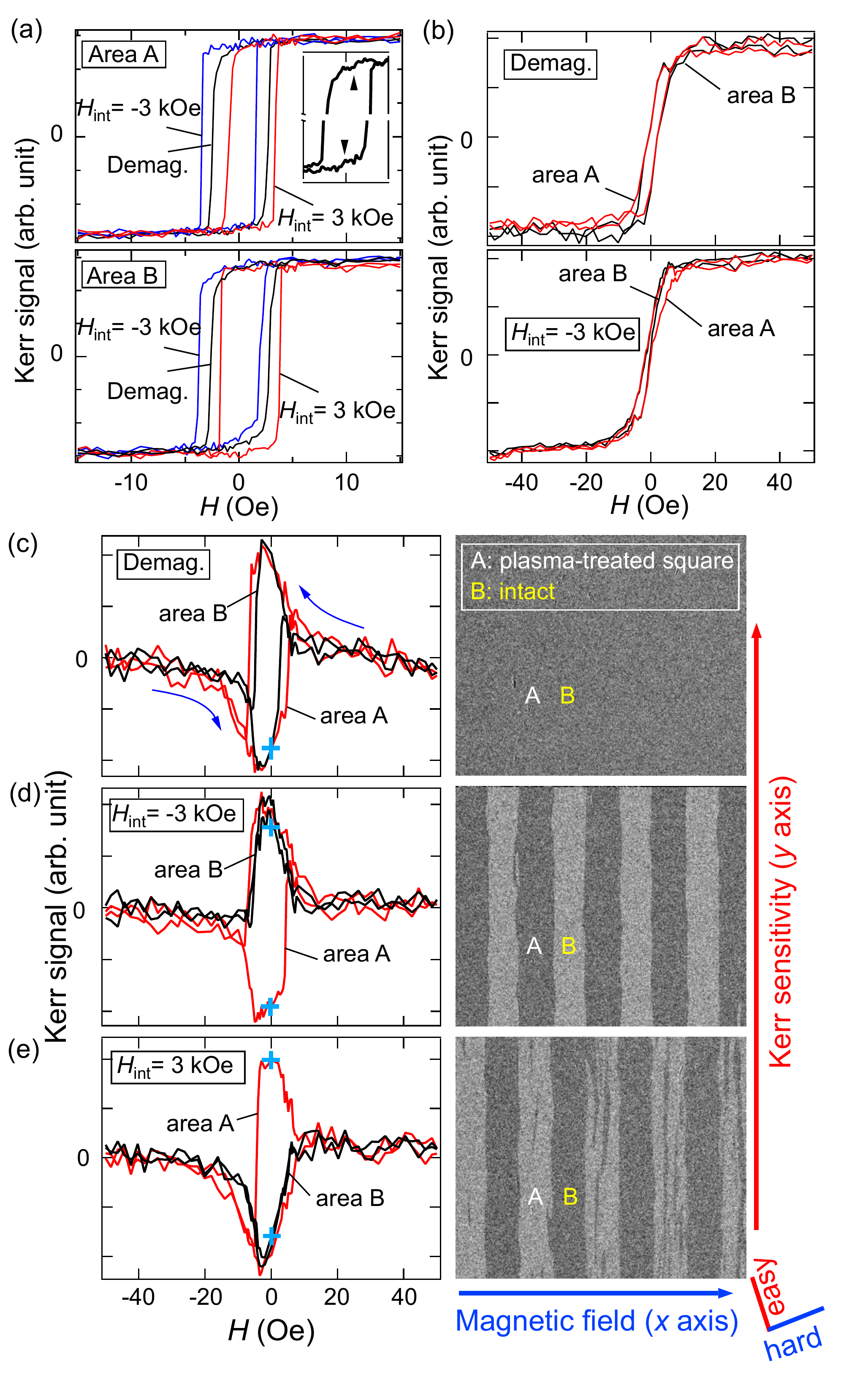}
	\caption{(Color online) Spatially-resolved MOKE hysteresis loops of areas A (plasma-treated) and B (intact) on the surface-patterned Fe oxide/FePd film (a) in the \textit{longitudinal} geometry with the external magnetic field along the direction close to the \textit{easy} axis, (b) in the \textit{longitudinal} geometry with the field direction close to the \textit{hard} axis, and (c)--(e) in the \textit{transverse} geometry with the field direction close to the \textit{hard} axis. Hysteresis loops are obtained with initial magnetization at $H_\mathrm{int}=\pm 3$ kOe along the direction close to the easy axis, and after demagnetization (labeled `Demag.'), respectively. The inset in (a) is a blown-up view of the top and bottom parts of the `Demag.' loop from area A. In (b)--(e), the loops of area A are red, and those of area B are black. In (c)--(e), arrows indicate the sweep directions, and blue crosses indicate the spots where the Kerr images on the right are taken.}\label{Fig6}
\end{figure}

To further investigate exchange interaction and anisotropy in our system, a high external magnetic field of $\pm$3 kOe is applied to the surface-patterned Fe oxide/FePd film along the direction (i.e., the $y$ axis) close to the easy axis to commence an initial magnetization, followed by measurements under small transient fields of 0 Oe $\rightarrow \pm 15$ Oe $\rightarrow$ 0 Oe along the direction close to the easy axis (Fig.~\ref{Fig6}(a)), or 0 Oe $\rightarrow \pm 50$ Oe $\rightarrow$ 0 Oe along the direction close to the hard axis (Figs.~\ref{Fig6}(b)--(e)). 

Fig.~\ref{Fig6}(a) displays the longitudinal MOKE hysteresis loops with the field direction close to the easy axis after initial magnetization.  The blue loops are obtained after initial magnetization at $-$3 kOe, the red ones are after initial magnetization at $+$3 kOe, and the black ones are after demagnetization following an initial magnetization at 3 kOe. The demagnetization is conducted using an oscillating magnetic field of decreasing magnitude until the magnetization is reduced to zero at zero field. The shifts of the loops with high-field initial magnetization with respect to the demagnetized one declare the existence of EB in our system, and that the directions of the EB fields can be freely controlled by the initial-magnetization directions. These loops shift in the same direction as the initial high field, which is referred as `positive' spontaneous EB. 

Setting EB usually involves cooling an AFM/FM bilayer below the N\'{e}el temperature in the presence of a magnetic field.\cite{Meiklejohn1957} It has also been shown that related phenomenon can occur in coupled two FMs with dissimilar magnetic properties,\cite{Fullerton1998,Berger2004} where the role of the AFM is fulfilled by a hard FM layer, and EB setting is realized by applying a sufficiently high initial magnetic field without the need of cooling preprocessing.\cite{Berger2004,Binek2006} This explains the EB effect observed in our system, where the Pd-rich layer with remnants of Fe is a soft FM\cite{Crangle1960}, whereas the FePd layer is an FM with weak PMA and an induced in-plane anisotropy near the interface (Fig.~\ref{Fig_interface}). The initial high magnetic field induces a non-zero net in-plane magnetic moment in the Pd-rich layer, which stems from an enlargement of the closure domains oriented parallel to the initial field at the expense of those oriented antiparallel to the field. The coupling between the Pd-rich layer and the net in-plane magnetization of the FePd layer then accounts for the observed shift of the hysteresis loop.\cite{Sort2004} The positive spontaneous EB arises from the interfacial AFM coupling between FePd and the Pd-rich layer, with a net interfacial moments that are not reversed during the measurements of the hysteresis loops.\cite{Won2007,Mangin2006}

It is worth noting that many of these hysteresis loops exhibit a pair of small kinks on the top and bottom of each hysteresis loop near the central field of the loop. An example is shown in the inset of the top panel of Fig.~\ref{Fig6}(a), which is blown-up view of the top and bottom parts of the `Demag.' loop in the panel, with the kinks indicated by the black arrows. These kinks are of magnitude close to the noise level and therefore can be easily missed, but their reproducibility confirms their existence. The pair of top and bottom small kinks points to an additional switching field attributed to a minority soft layer in the system,\cite{Fullerton1998,Chen2019} which in our case is the Pd-rich layer. These kinks have also been observed in the hysteresis loop of a SmCo film with a mixture of Sm$_x$Co$_y$ phases being present,\cite{Fullerton1998} which play similar roles to the FePd phase and the Pd-rich phase in our system.  

It is noticed in Fig.~\ref{Fig6}(a) that the loops taken inside (labeled `area A') and outside (labeled `area B') the $50 \times 50$ $\upmu$m$^2$ square exhibit positive EB of similar magnitudes. Since the Fe oxide in area B was naturally formed in ambient air without plasma treatment, the Pd-rich layer under the oxide is not as `rich' as that in the plasma-treated area A, but seems to be able to generate EB of a similar magnitude to that of area A, at least under the condition of initial magnetization at $\pm$3 kOe. 
However, the EB effect in area A can be easily distinguished from that in area B if longitudinal MOKE hysteresis loops are recorded with the magnetic field applied along the direction close to the \textit{hard} axis, as shown in Fig.~\ref{Fig6}(b). With initial magnetization at $-$3 kOe, the loop of area B (black curve) is apparently slimmer than that of area A (red curve) in the positive signal, which reveals different anisotropy between the two areas.

The interplay between the EB and the anisotropy further manifests itself in the spatially-resolved \textit{transverse} MOKE hysteresis loops shown in Figs.~\ref{Fig6}(c)--(e) with the field direction close to the hard axis. Fig.~\ref{Fig6}(c) shows the loops taken after demagnetization, accompanied with a corresponding Kerr image taken at 0 Oe. The two loops from areas A and B approximately overlap with each other, and thus the stripe domain pattern is barely seen in the Kerr image. However, if the sample is measured after initial magnetization at a high field without demagnetization, distinct behaviors are observed as displayed in Figs.~\ref{Fig6}(d) and \ref{Fig6}(e). Whereas the loop of the intact area B approximately follows the route seen in Fig.~\ref{Fig6}(c), the loop of the plasma-treated area A stays in the same sign of the magnetization along the Kerr-sensitivity direction, which is positive with initial magnetization at a field of $-$3 kOe (Fig.~\ref{Fig6}(d)), and negative with initial magnetization at $+$3 kOe (Fig.~\ref{Fig6}(e)). This again implies an in-plane AFM coupling existing in the system. The opposite magnetization in areas A and B as the field is swept to 0 Oe at the spots marked with the blue crosses on the hysteresis loops leads to magnetic-domain patterns with clear contrast shown in the Kerr images on the right. This clear contrast at zero field cannot be achieved without the presence of EB. Besides, the stripe domains with EB (Kerr images in Figs.~\ref{Fig6}(d) and \ref{Fig6}(e)) are straighter than the ones observed without EB (lower panel in Fig.~\ref{Fig2}(c)). 

The phenomenon observed in Figs.~\ref{Fig6}(c)--\ref{Fig6}(e) may be understood in terms of the competition between the UMA of our sample (i.e., having one easy axis) and the EB set by the high initial external field applied along the $y$ axis, which is at an angle of $\sim$15$^{\circ}$ to the easy axis. When no EB is present, as in the case after demagnetization displayed in Fig.~\ref{Fig6}(c), the hysteresis loops are similar to what we have observed in Fig.~\ref{Fig2}(b). However, as EB becomes present, one needs to consider the interplay between EB and UMA. According to theoretical studies,\cite{Meiklejohn1957, Chung2005} the look of the transverse MOKE hysteresis loop with the field applied orthogonal to the EB direction depends on the relative strength of UMA to EB and the angle between their directions. UMA tends to exert a torque to rotate the moments, and hence should result in opposite signs of the magnetization along the direction orthogonal to the external field when EB is small, provided that the angle between the directions of UMA and EB is large enough ($> \sim$15$^{\circ}$). If the angle between them is too small, or if EB is strong compared to UMA, the direction of the magnetization is determined dominantly by EB instead, leading to the same sign of the magnetization along the Kerr-sensitivity direction (i.e., parallel or antiparallel to EB) during both reversals. For the intact area B, the Kerr signal has the same sign during both reversals of its transverse MOKE hysteresis loop, which is positive in Fig.~\ref{Fig6}(d)  and negative in \ref{Fig6}(e). This is probably because the angle between the UMA (i.e., the easy axis) and the EB (set by the initial high field) is small ($\sim$15$^{\circ}$). For the plasma-treated area A, however, each transverse MOKE hysteresis loop exhibits both positive and negative branches. This may be explained by the new UMA created after plasma treatment, as implied in Figs.~\ref{Fig3}(c) and \ref{Fig3}(d), which has a direction at a much larger angle of $\sim$45$^{\circ}$ to the EB. The phenomenon of the interplay between EB and UMA further confirms the effectiveness of our microscopic magnetic-property modulation method that conveniently utilizes lithographed plasma treatment on the Fe oxide/FePd thin films. The drastic difference in the microscopic magnetization behavior between the two cases with and without EB permits a practical switch from one mode of magnetic-domain behavior to another by simply turing EB on or off at room temperature.

\section{Conclusions}

In summary, we demonstrate control of magnetic domain structures of magnetically micro-patterned Fe oxide/FePd thin films through e-beam lithography and O$_2$- or Ar-plasma treatment of the surface. Transverse MOKE measurements with the magnetic field applied along the direction close to the magnetic hard axis reveal that the magnetic field required to reverse the magnetization of the plasma-treated areas inside the $50 \times 50$ $\upmu$m$^2$ squares in periodic arrays is larger than that for the untreated areas, except that the untreated areas that are aligned with the treated squares along the easy axis are magnetically coupled with the squares. This results in periodic stripes of domains observed in the transverse MOKE images during magnetization reversal. Besides, an intriguing competition between the uniaxial anisotropy and the exchange bias is observed in our system, and can be microscopically controlled by altering the magnetic anisotropy via lithographed plasma treatment. The method of surface patterning with plasma treatment for magnetic-domain engineering may be applied in design and fabrication of future data-storage and spintronic devices.

\section*{Acknowledgment}

We acknowledge the group of Prof. Hsiang-Chih Chiu for assisting us with the plasma source. This study is sponsored by the Ministry of Science and Technology of Taiwan under Grants Nos.~MOST 105-2628-M-003-001-MY3, MOST 105-2633-M-003-001, and MOST 107-2112-M-003-004.

\section*{References}
\bibliography{FePd_2}
\bibliographystyle{elsarticle-num}

\newpage

\end{document}